\documentclass[a4paper,11pt]{article}
\usepackage{jcappub} 
\usepackage[normalem]{ulem}
\usepackage{aas_macros}

\usepackage{xcolor}

\usepackage[style=numeric,backend=biber, sorting = none]{biblatex}

\arxivnumber{7777.77777} 
\title{\boldmath LOO-PIT: A sensitive posterior test}

\author[a,b]{Alan B. H. Nguyen}
\author[a,b]{Marco Bonici}
\author[c]{Glen McGee}
\author[a,b,d]{Will J. Percival}

\affiliation[a]{Waterloo Centre for Astrophysics, University of Waterloo, 200 University Ave W, Waterloo, ON N2L 3G1, Canada}
\affiliation[b]{Department of Physics and Astronomy, University of Waterloo, 200 University Ave W, Waterloo, ON N2L 3G1, Canada}
\affiliation[c]{Department of Statistics and Actuarial Sciences, University of Waterloo,
Waterloo, ON N2L 3G1, Canada}
\affiliation[d]{Perimeter Institute for Theoretical Physics,
31 Caroline St. North, Waterloo, ON N2L 2Y5, Canada}

\emailAdd{alan.nguyen@yale.edu}

 \abstract{With the advent of the next generation of astrophysics experiments, the volume of data available to researchers will be greater than ever. As these projects will significantly drive down statistical uncertainties in measurements, it is crucial to develop novel tools to assess the ability of our models to fit these data within the specified errors. We introduce to astronomy the Leave One Out-Probability Integral Transform (LOO-PIT) technique. This first estimates the LOO posterior predictive distributions based on the model and likelihood distribution specified, then evaluates the quality of the match between the model and data by applying the PIT to each estimated distribution and data point, outputting a LOO-PIT distribution. Deviations between this output distribution and that expected can be characterised visually and with a standard Kolmogorov--Smirnov distribution test. We compare LOO-PIT and the more common $\chi^2$ test using both a simplified model and a more realistic astrophysics problem, where we consider fitting Baryon Acoustic Oscillations in galaxy survey data with contamination from emission line interlopers. LOO-PIT and $\chi^2$ tend to find different signals from the contaminants, and using these tests in conjunction increases the statistical power compared to using either test alone. We also show that LOO-PIT outperforms $\chi^2$ in certain realistic test cases. }

\begin{document}
\maketitle
\flushbottom

\section{Introduction}

With the next generation of galaxy surveys and cosmological experiments on the horizon, including the Roman Space Telescope \cite{spergel2013widefield}, Euclid \cite{laureijs2011euclid, mellier_euclid}, DESI \cite{desi_survey_design}, LSST \cite{lsst_rubin_design}, and CMB-S4 \cite{cmbs4_design}, we will soon have unprecedentedly precise constraints on cosmological models. The accuracy with which we can create models to fit to the data, including likelihoods and potential systematic problems, needs to improve such that it is not the limiting factor in the inferences we can make from the upcoming data. It is also crucial to develop techniques to assess the accuracy of these inferences. The astrophysical community has traditionally relied on tests like the $\chi^2$. While it is indeed a simple and powerful test whose properties are well understood, it has some limitations and is not able to detect all the possible issues within an inference pipeline. For this reason, more advanced statistical techniques are becoming increasingly common in cosmological analyses for posterior checks \cite{feeney2019, desy1, rogers_peiris2021}. These include the Akaike Information Criterion (AIC), as well as recent methods that quantify tensions, check the consistency of data sets, or include the model selection as a part of the analysis \cite{bma_2024, tension_metrics_2021, concordance_2019}.

In this paper we introduce a new technique to astronomy, called the Leave One Out-Probability Integral Transform (LOO-PIT). LOO-PIT has previously been applied in statistics \cite{firstpit, pit_uniformity}, and we now demonstrate how it can be applied to cosmological problems. LOO-PIT combines two concepts: Leave One Out-Cross Validation (LOO-CV), and the Probability Integral Transform (PIT) \cite{Vehtari_Gelman_Gabry_2016}. 

Leave One Out-Cross Validation (LOO-CV) and the Probability Integral Transform (PIT) are both important techniques used in statistical model evaluation. LOO-CV is a method that assesses a model's performance by removing one data point at a time from the dataset, training the model on the remaining data, and predicting the excluded point. By repeating this process for every data point, LOO provides a robust estimate of a model's predictive performance across the data set.

In contrast, the Probability Integral Transform (PIT) evaluates how well a probabilistic model fits the data. The PIT transforms the observations using the model's cumulative distribution function (CDF), resulting in values that, for a well-calibrated model, should be uniformly distributed between 0 and 1. The PIT is particularly effective in checking whether the model's predicted probabilities align well with the actual outcomes.

LOO-PIT brings these two methods together, offering a powerful technique combining the strengths of each. In LOO-PIT, we first use the model to predict the distribution of each excluded data point. Then the CDF of the predicted distribution is evaluated at the observed value of the left-out data point, applying the PIT to transform the prediction into a value between 0 and 1. If the model is well-calibrated, the resulting LOO-PIT values will be uniformly distributed. LOO-PIT can not only measure the accuracy of the model’s predictions but also assess how well the model captures uncertainty in those predictions. Any deviation from uniformity indicates modelling failure, and the shape of the LOO-PIT distribution can be used to indicate which type. As a result, LOO-PIT can be widely used in Bayesian modelling, where proper uncertainty quantification is critical, ensuring that both the point predictions and the uncertainty estimates of a model are trustworthy.

A complementary tool to the standard $\chi^2$ goodness of fit test, LOO-PIT provides both a diagnostic plot characterising the type of modelling failure and a framework for testing model fit by comparing distributions via the Kolmogorov--Smirnov test (and others). LOO-PIT also relaxes the constraint of using Gaussian statistics when assessing the quality of the modelling, which are implicitly assumed when using the $\chi^2$ statistic. Furthermore, $\chi^2$ uses the information from a single point in the parameter space, while LOO-PIT incorporates information from the entire posterior distribution, leading to a better global assessment of the inference.

To demonstrate how LOO-PIT can be used to diagnose modelling failures, we apply the LOO-PIT technique to the problem of emission line interlopers in galaxy clustering data. Galaxy catalogues can be contaminated by sources where the redshift has been incorrectly inferred due to emission line confusion. As the distance to a galaxy is derived from its redshift, these interlopers contaminate the clustering signal and in particular can offset the BAO peak. We fit contaminated correlation functions with a standard BAO template and apply LOO-PIT to the posteriors obtained to determine its sensitivity to both the interloper fraction and the interloper displacement.

Our paper is structured as follows. In Section~\ref{sec:method_outline} we present the LOO-PIT technique, and demonstrate it on a simple model in Section~\ref{sec:simp_model}, highlighting common LOO-PIT detection signatures. In Section~\ref{sec:loopit_bao}, we briefly discuss the effects of small displacement interlopers on BAO measurements, and apply LOO-PIT to test the goodness of fit in the presence of these interlopers. Some limitations of the method are discussed in Section~\ref{sec:baseline_calibration}, focusing on its application to astronomical data. Finally, we conclude in Section~\ref{sec:conclusions}.

\section{Method Outline}
\label{sec:method_outline}
This method combines two statistical techniques---Leave One Out-Cross Validation (LOO-CV) and the Probability Integral Transform (PIT)---to form a sensitive posterior check.  A partial implementation of the method presented in this paper can be found using the \texttt{arviz} package \cite{arviz_2019}. In this section, we provide an overview of the LOO and the PIT parts of this technique and compare to the commonly used $\chi^2$ statistic.

\subsection{Outlining LOO-PIT}

Generally, when fitting data with a model, we are interested in how likely the data is, given the model, or how likely the model is, given the data. These are related to the likelihood and the posterior, respectively. LOO-PIT and other posterior checks we are interested in a slightly different question: under the assumption of some model, how likely is some subset of our data given the remainder? In other words, does our model predict the data well? It is this question that the LOO-PIT procedure attempts to answer. 

We start by estimating a \textit{posterior predictive distribution}. This is the distribution of possible future observations (or unobserved data) based on the data one has already observed and the model's posterior distribution of the parameters. It represents the uncertainty in predicting new data by integrating over the uncertainty in the model's parameters, reflecting both the variability in the data and the uncertainty in parameter estimates. In astronomy, where future data is often lacking, LOO offers a practical method to simulate the outcomes we might observe if we did possess future data.

Let $y_i$ be the ``left out'' observation, $y_{-i}$ the remaining observations, and $\theta$ the model parameters. The posterior predictive distribution is then defined as:
\begin{equation}
    p(y_i|y_{-i}) = \int p(y_i|\theta)p(\theta|y_{-i})d\theta.
    \label{eq:ppd}
\end{equation}
The observed data points $y$ can be considered to be drawn from some unknown distribution. If we specify the correct model for the observed data, the posterior predictive distributions should match this underlying distribution. That is, the model correctly predicts our data. We compute the posterior predictive for $y_{i}$, the ``left out'' observation, given the remaining observations $y_{-i}$. We then repeat this for each of $i=1,\dots,n$ observations, providing an estimate for that unknown distribution of each observation.

If the model is correctly specified, the posterior predictive distribution for each left out point should match their true distribution. Thus transforming an observation using the estimated CDF from the posterior predictive would be the same as transforming with the true underlying distribution and should yield a single draw from a uniform distribution. This is known as the Probability Integral Transform (PIT). Repeating this process for each observation then gives a sample of LOO-PIT values, which we can compare to its expected distribution under correct model specification. While this expectation is a uniform distribution under a well calibrated model specification, there are situations where, by construction, even the correct model does not result in uniform draws when applying LOO-PIT. In these cases, we must first characterise the baseline LOO-PIT distribution under correct model specification. 

Once the baseline is specified, we then consider two related diagnostics to compare the LOO-PIT distributions to the expectation. First, is simply the kernel density estimate (KDE) of all the estimates together. It is straightforward to check if it is consistent with a uniform (or some baseline) by inspection, in addition to determining the type of modelling failure based on how the LOO-PIT KDE deviates from the uniform. We can also test the uniformity of the LOO-PIT KDE using the Kolmogorov--Smirnov (KS) Test (or another distribution test), yielding $p$-values for the null hypothesis.

\subsection{Estimating Leave-One-Out Posterior Predictives}

The first step of LOO-PIT is to estimate the posterior predictive distribution in Equation~\ref{eq:ppd}.
While one could perform exact LOO-CV by obtaining the $p(\theta|y_{-i})$ posteriors via MCMC, it would be cost prohibitive to refit the model $N$ times to obtain posteriors  for each LOO subset. Instead, we use importance sampling (IS) to obtain these probabilities from a single model fit. IS is a Monte Carlo technique used to estimate properties of a target distribution by drawing samples from another distribution and then weighting them to correct for the difference between the two distributions. For a more in-depth review, see \cite{importance_sampling_review}.

By \cite{Vehtari_Gelman_Gabry_2016}, if the pointwise probabilities are conditionally independent given the parameter values,\footnote{One can decompose $p(y|\theta)$ = $\Pi^{N}_{i}p(y_i|\theta)$.} the importance sampling estimate for the posterior predictive for the $i^{th}$ observation is:
\begin{equation}
    \hat{p}(y_{i}|y_{-i}) = \sum^{S}_{s=1} r_{i}^{s} \ p(y_{i}|\theta^s).
    \label{eq:pointwise_IS}
\end{equation}

In general, the likelihood is not fully factorisable. However, for models where the data is multivariate Gaussian distributed with a non-diagonal covariance matrix, there is a simple modification to the above equation \cite{multigauss_prob_deriv}:\footnote{See \cite{multigauss_prob} for the exact form of $p(y_{i}|y_{-i}, \theta^s)$.}

\begin{equation}
    \hat{p}(y_{i}|y_{-i}) = \sum^{S}_{s=1} r_{i}^{s} \ p(y_{i}|y_{-i}, \theta^s),
    \label{eq:pointwise_IS_multigauss}
\end{equation}
\noindent
where $r_{i}^{s}$ are the standard normalised importance ratios and $S$ is the total number of parameter samples. The importance ratios are given by:\footnote{In this equation, $p(\ldots)$, refers to the posterior probability of the given draw from $p(\ldots)$}
\begin{equation}
    r_{i}^{s} = \frac{p(\theta^s|y_{-i})}{p(\theta^s|y)}.
\end{equation}

In cases where the two distributions, $p(\theta|y_{-i})$ and $p(\theta|y)$, differ substantially, the distribution of the IS ratios can have a heavy right tail. The IS estimate can be dominated by these extreme samples and give unstable estimates. There are two commonly applied ``smoothing" algorithms that mitigate this issue with standard IS. The first is Truncated Importance Sampling (TIS) introduced by \cite{ionides2008TIS}. TIS replaces the raw IS ratios with truncated ratios, $t^s$, given by:
\begin{equation}
    t^s_{i} = {\rm min}(r^s_{i}, \sqrt{S}\bar{r_{i}}),
\end{equation}
\noindent
where $\bar{r_{i}}$ is the sample mean of the raw IS ratios. While this method is simple to implement and guarantees that  estimates have finite variance, TIS introduces a potentially large bias \cite{Vehtari_Gelman_Gabry_2016}.

Instead, we adopt a Pareto Smoothed Importance Sampling (PSIS) scheme \cite{vehtari2016PSIS}. PSIS is similar to TIS in that we replace some of the raw IS ratios, though it does not rely on a simple truncation. First, the raw IS ratios are ordered from least to greatest and indexed $s$ = 1, 2, \dots, $S$. The top $M$ are then fit to a generalised Pareto distribution using the method of \cite{zhang_stevens_2009_paretofit}.\footnote{$p(x|\mu,\sigma,k) = \frac{1}{\sigma}\left(1 + k\left(\frac{x-\mu}{\sigma}\right)\right)^{-\frac{1}{k} - 1}$} Then for a given point indexed by $i$, the raw IS ratios are replaced as:
\begin{align}
    w^{s}_{i} = r^s_i, &\quad s = 1, 2, \dots, S - M \\[0.7em]
    w^{S - M + z}_{i} = {\rm min}\left(F^{-1}\left(\frac{z - 1/2}{M}\right), {\max }(r^{s}_{i})\right), &\quad z = 1, 2, \dots, M,
\end{align}

\noindent
where $F^{-1}(\ldots)$ is the inverse CDF of the generalized Pareto distribution, and $M$ is the number of samples used for the fit. By default we set $M = 0.2S$.\footnote{Note that the method is robust to the mis-specification of the exact form of M relative to S \cite{vehtari2016PSIS}.} The final step is to truncate the smoothed ratios as in TIS, but using the cutoff of $S^{3/4}\bar{w_{i}}$, to guarantee finite variance in the estimates.

Finally, with the smoothed ratios, we have an estimate for each posterior predictive distribution using LOO-PSIS:
\begin{equation}
    \hat{p}(y_{i}|y_{-i}) = \sum^{S}_{s=1} w_{i}^{s} \ p(y_i|y_{-i}, \theta^s)\,.
    \label{eq:pointwise_PSIS_pp}
\end{equation}

This is our estimate of the distribution of a single point $y_i$ given the other data points $y_{-i}$, marginalised over the parameter space; we refer to this as the \emph{pointwise posterior predictive distribution}.

IS, or its refined forms, is not necessary to perform the LOO-PIT procedure. It is only to reduce the computational burden. If one were keen to avoid the approximations that IS and PSIS introduce, it is possible to rerun the chains for each left out data points to get the exact leave one out posteriors. While this procedure is exact, it significantly increases the computation time as the number of data points increases. Using the IS approximations allows LOO-PIT to scale to larger  datasets without unnecessarily long runtimes. However, there may be cases where PSIS fails due to the nature of the posteriors. In these cases, then it would be necessary to resample the posteriors, but these cases are likely rare.

Before we can apply the PIT in the next step, we must estimate the pointwise posterior predictive CDF. The posterior predictives are t-distributed with large degrees of freedom, related to the number of MCMC draws, and are well approximated by a Gaussian \cite{johnson_geisser1983_tdist_pp}. Thus, we approximate these distributions to a Gaussian with a mean and variance that match those of the set of $y^s_i$ weighted by the PSIS ratios $w^s_i$.

\subsection{The Probability Integral Transform}
\label{sec:pit}

To test the posterior predictive distributions, we transform each observation with its corresponding pointwise posterior predictive CDF. This is where we apply the Probability Integral Transform.

Consider a continuous random variable, $X$, with PDF $p(X)$ and CDF, $F(X)$. Then the random variable $Y$, is defined by the following transformation of $X$:
\begin{equation}
    Y(X) = \int^{X}_{-\infty} p(X^*) dX^*\,,
    \label{eq:defY}
\end{equation}

\noindent
that is, $Y=F(X)$. We can find the PDF of $Y$, $p(Y)$, in terms of $p(X)$ by the change in the PDF under a change of variables (because $F$ is monotonically increasing):
\begin{equation}
    p(Y) = \left|\frac{dX}{dY}\right|p(X).
\end{equation}
\noindent
Combining this with Equation~\ref{eq:defY}, we find:
\begin{equation}
    p(Y) = \left|\frac{1}{p(X)}\right|p(X) = 1\,.
\end{equation}

\noindent
That is, $Y$ is uniformly distributed.

If, by contrast, the incorrect CDF is considered, the Jacobian will not result in 1/$p(X)$, and thus the distribution of $Y$, $p(Y)$, will deviate from the uniform. The test is then to determine whether the distribution of these LOO-PIT values is consistent with a uniform distribution.

Using this CDF to transform each of the $N$ observed data points $y^{\rm obs}_{i}$ will result in $N$ LOO-PIT values. The value of the $i^{\rm th}$ LOO-PIT draw is given by:

\begin{equation}
    L_{i} = \int^{y^{\rm obs}_{i}}_{-\infty} p(X^*) dX^*\,,
    \label{eq:Lval_transform}
\end{equation}
\noindent
where $p(X^*)$ is the corresponding $i^{\rm th}$ pointwise posterior predictive distribution $\hat{p}(y_{i}|y_{-i})$. Combining these LOO-PIT draws together forms the KDE that we can inspect for signs of modelling failures.

\subsection{\texorpdfstring{Conceptual differences from the $\chi^2$ test}{Conceptual differences from the X2 test}}

The standard $\chi^2$ goodness of fit statistic is the sum of the squares of $k$ independent standard \textit{Gaussian} random variables. In general, these are the residuals of the model: $data - model$. When these aresian distributed, the $\chi^2$ test is valid. However, in cases where this underlying assumption is invalid, such as with skewed residuals, $\chi^2$ is not valid.

LOO-PIT has two primary advantages over $\chi^2$. The most general advantage is that LOO-PIT does not require that one assumes that the true underlying distribution of the data is some multivariate Gaussian. In our analyses, the Gaussian CDF is merely a computationally efficient approach that approximates the true posterior predictive well and allows for the analytic formulae for the leave one out posterior predictives. We also assume the data are Gaussian distributed so that the $\chi^2$ test is also a valid test for comparison purposes.

While $\chi^2$ quantifies the quality of the modelling, it is not sensitive to a common type of modelling error: bias. This is where LOO-PIT can succeed where $\chi^2$ fails: by incorporating information about bias in the model, LOO-PIT contains more information than a single $\chi^2$ statistic. Indeed, the shape of the LOO-PIT diagnostic plot signals the type of model failure, expediting the formulation of better models. Furthermore, as $\chi^2$ is a sum of squared differences, it tends to be stronger in detecting outliers. LOO-PIT, with its additional information about the direction of small biases, can best detect these small correlated offsets. As such, using LOO-PIT and $\chi^2$ in tandem can help detect a range of different types of model failure. We will demonstrate this in Section~\ref{sec:large_displacement_loopit}.

\subsection{Summary of Method}

\begin{enumerate}
    \item \textit{Characterise the baseline LOO-PIT distribution}\\
    See Section~\ref{sec:baseline_calibration} for a detailed discussion of this point.

    For cosmological data, we can use simulated data to characterise a baseline without any contaminants. Thus, if we detect issues with LOO-PIT, we would know we have deviations from the fiducial assumptions.
    
    \item \textit{Run MCMC chains}\\
    This is the standard modelling step.

    \item \textit{Apply LOO-PIT}\\
    Armed with the MCMC chains, we can estimate the posterior predictive distributions with approximate or exact LOO, then apply the PIT to each data point and estimated posterior predictive distribution, combining into a KDE to obtain a LOO-PIT distribution.

    \item \textit{Interpret results}\\
    In general, the resulting LOO-PIT distribution is likely to be some combination of the basic shapes shown in Section~\ref{sec:common_signatures}. Furthermore, we can estimate the significance of the detection by combining LOO-PIT with either the KS test or another distribution test.
\end{enumerate}

\section{Testing LOO-PIT in a Simple Model}
\label{sec:simp_model}

In this section, we demonstrate three basic LOO-PIT signatures and test LOO-PIT on a simple model without the need to perform the baseline calibration step.

\subsection{Common LOO-PIT signatures}
\label{sec:common_signatures}

\begin{figure}[htbp]
\centering
\includegraphics[width=\textwidth]{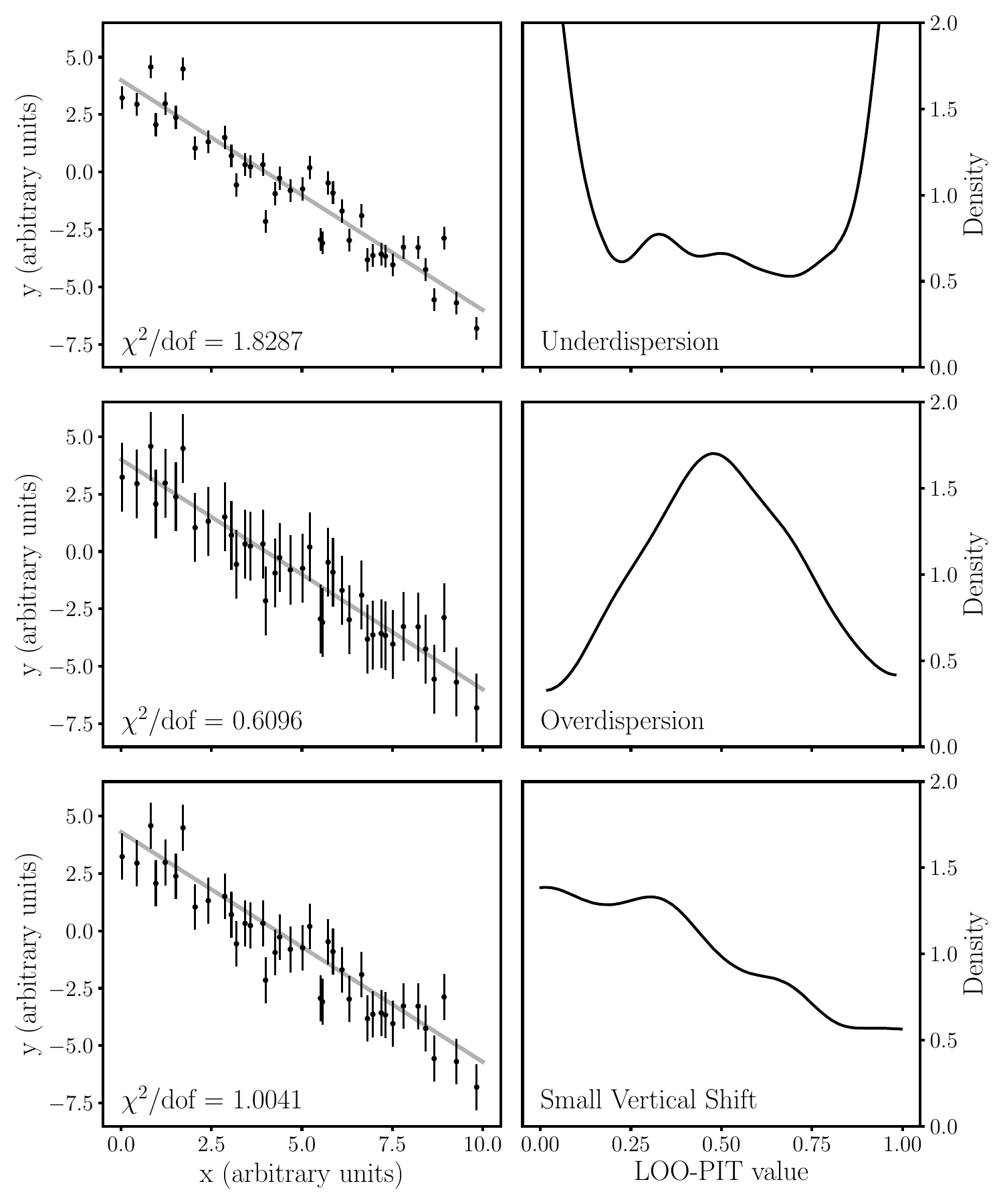}
\caption{LOO-PIT values for a simple linear fit on 200 points. The data is thinned in the left columns plots for clarity. In the first row, the error bars are 0.5x their true values, showing the underdispersion ``trough'' signature. In the second row, the error bars are 1.5x their true values, showing the overdispersion ``peak'' signature. In the third row, the plot y values are offset by -0.5, showing the bias ``tilt'' signature. \label{fig:example}}
\end{figure}

We show three common LOO-PIT signatures of model failure in Figure~\ref{fig:example}, which is based on \cite{Pla}. These are underdispersed/underfitting, overdispersed/overfitting, and biased/shifted. 

The first two effects are related to the estimation of the covariance matrix of the data. As the covariance is used to calculate the posteriors and likelihoods which are then applied to estimate the posterior predictive distributions, improper estimation of the covariance matrix is detectable by LOO-PIT. For data where the covariances are underestimated, the posterior predictive distributions will also be narrower than the truth. Thus, data points will tend to fall on the tails of the posterior predictive distributions, forming the trough pattern seen in the first row of Figure~\ref{fig:example}. This pattern can also appear if the model is not a good fit for the data, such as fitting quadratic data with a linear model. Since the posterior predictives are based on the model, if the model and data are significantly different, the LOO-PIT values will again be drawn from the tails of the posterior predictives. Imagine fitting quadratic data with a linear model. While the posterior predictives would be closely centred on the best fit line, the quadratic data will fall into the tails of these distributions. 

The opposite effect occurs when the covariances are over-estimated. Now the posterior predictives will be too wide, and data points will tend to fall close to the middle of these distributions. Thus, the LOO-PIT draws are close to 0.5 and we will get the peak pattern shown in the second row of Figure~\ref{fig:example}. This pattern also appears when the model is overfitting. Suppose the 10 points of data are fit with a 10 parameter model. The model can perfectly fit each data point, thus the LOO-PIT draw for each point will be exactly 0.5. These two signatures can either be caused by poor covariance estimation or poor model specification. It is then important to correctly estimate errors and covariances, lest a true model mis-specification be attributed to poor error estimation.

The final common signature is a shift or bias in the data. This effect is also what provides LOO-PIT additional information compared to $\chi^2$. Depending on the direction of the bias, the LOO-PIT KDE can have either a positive or negative tilt. This is shown in the third row of Figure~\ref{fig:example}. 

\subsection{Goodness of Fit Tests}

We now test the sensitivity of LOO-PIT+KS on a simplified toy model. We start with a well-calibrated baseline model for linear data that returns the expected uniform LOO-PIT shape. We then add a Gaussian feature to the data, and increase its amplitude as a proxy for increasing levels of contamination.

\begin{figure}[htbp]
\centering
\includegraphics[width=\textwidth]{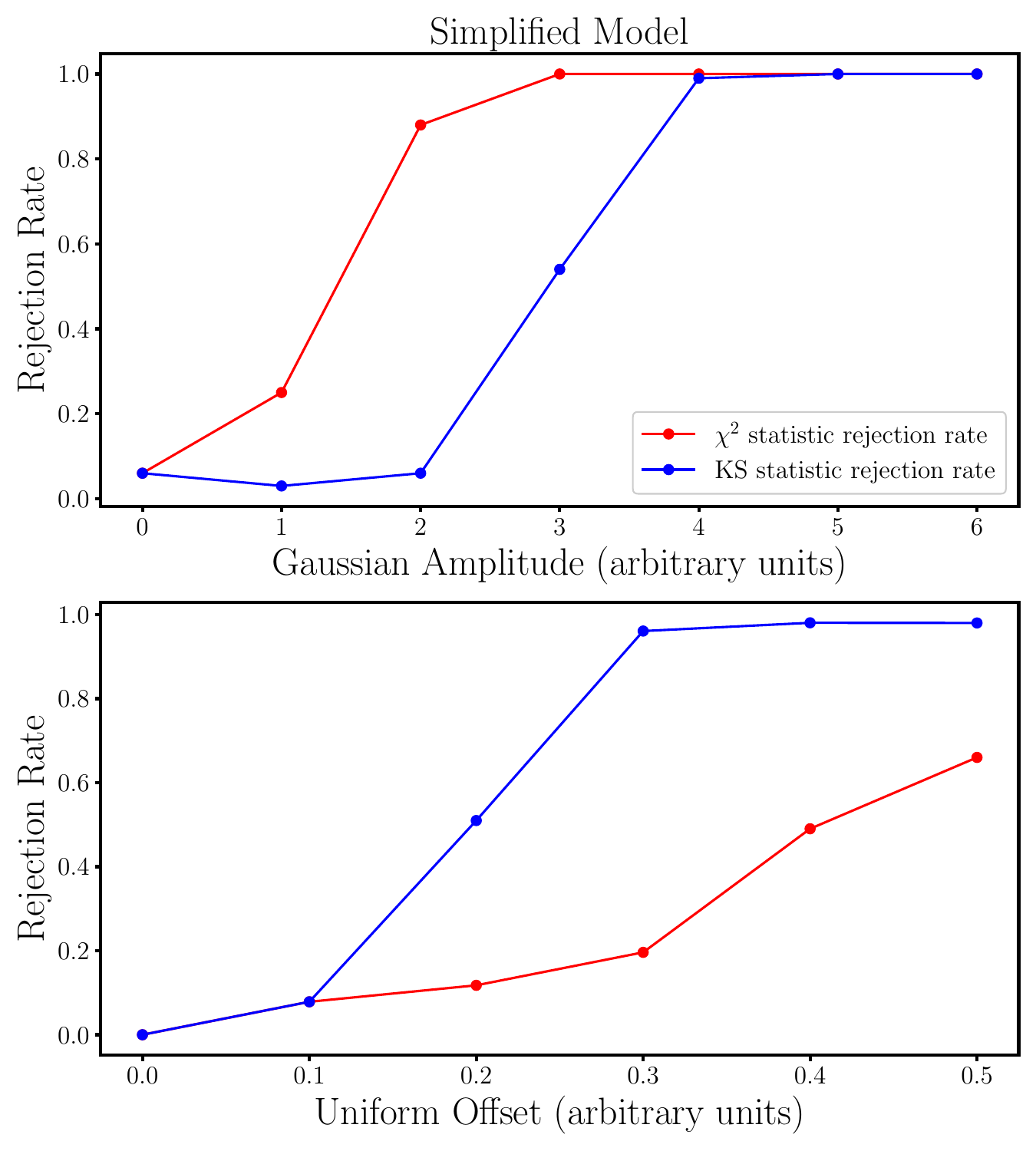}
\caption{Power plots comparing the rejection rates of $\chi^2$ and LOO-PIT+KS. In the upper panel, we applied both tests to a linear toy model with a Gaussian feature with increasing amplitude. In the lower panel we applied both tests to a linear toy model with increasing uniform biases. $\chi^2$ is shown in red and LOO-PIT+KS is shown in blue. \label{fig:toy_model}}
\end{figure}

We began by generating 100 realisations of linear data with no Gaussian feature. We then apply the LOO-PIT procedure to the MCMC chains from each realisation, resulting in 100 LOO-PIT distributions. As expected, these LOO-PIT distributions are consistent with the uniform distribution. The KS statistic also returns the correct Type I error rate of five percent.\footnote{Assuming a significance level of $\alpha$ = 0.05 and a null hypothesis of zero contamination.} We then increase the level of contamination by adjusting the height of the Gaussian feature from 0 to 6 in integer steps. We calculate the $\chi^2$ and KS statistic and compare their power to reject the null hypothesis. We find that $\chi^2$ is a more powerful statistic compared to LOO-PIT+KS for this type of contamination. The rejections rates are compared in the upper panel of Figure~\ref{fig:toy_model} for the two tests. 

Both tests are well calibrated with the nominal Type I error rate at zero contamination. However, $\chi^2$ begins to detect the contamination immediately at an amplitude of 1 and becomes saturated at an amplitude of 3. However, LOO-PIT+KS starts to detect at a significant rate at an amplitude of 3. Both tests become saturated at near 100 percent rejection at amplitude of 4.

As $\chi^2$ is based on the sum of square differences between the model and data, it is sensitive to outliers such as the contamination added in the above example. Conversely, LOO-PIT is sensitive to more systematic offsets, such as a uniform bias across the data. This scenario can better exploit the directional information that LOO-PIT incorporates into its test. We demonstrate this effect by comparing MCMC chains of the calibrated linear model to data with an increasing uniform bias. In this case, LOO-PIT+KS significantly outperforms $\chi^2$ even at small bias amplitudes. We show this in the lower panel of Figure~\ref{fig:toy_model}.

We now continue on with LOO-PIT by testing with a BAO template fit to catalogues with increasing levels of interloper contamination.

\section{Testing LOO-PIT on BAO fitting in the presence of interlopers}
\label{sec:loopit_bao}

In this section, we briefly discuss the effects of emission line interlopers on the BAO fit and the standard BAO template model. We test LOO-PIT on these contaminated catalogues and compare its statistical power against $\chi^2$ across several combinations of interloper fraction and interloper displacement.

We present the BAO model in Section~\ref{sec:bao_model}, and briefly outline the issue of small displacement emission line interlopers in Section~\ref{sec:interlopers}. See \cite{nguyen2024, foroozan2022, massara2023, massara2020} for a more detailed explanation of the effects of this systematic.
The results are presented in Section~\ref{sec:bao_results}.

\subsection{The BAO model}  \label{sec:bao_model}

The standard BAO template fit is designed to measure the projected position of the BAO, quantified by two parameters, $\alpha$ and $\epsilon$, better known as the Alcock-Paczynski (AP) parameters \cite{ap}. These measure the deviation of the BAO scale from that in a fiducial cosmological model.

The model fitted to the data uses nuisance parameters to marginalise over the non-BAO clustering signal, and models of the following form are typically used \cite{Ross-BOSS}:
\begin{equation}
    \xi_{\text{model, $\ell$}}(a_1,a_2,a_3,B,\alpha, \epsilon, r) = A_{\ell}(r) + B\times \xi_{\text{galaxy, $\ell$}}(\alpha, \epsilon, r)\,.
    \label{eq:baotemplate}
\end{equation}
Here $\xi_{\rm galaxy}$ is the large scale auto correlation function template with the multiplicative factor $B$ added to adjust the amplitude. $A(r)$ is a polynomial term of the form:
\begin{equation}
    A_{\ell}(r) = \frac{a_{1,\ell}}{r^{2}} + \frac{a_{2, \ell}}{r} + a_{3,\ell}\,.
\end{equation}

Rather than fit to the full clustering signal, we compress into monopole and quadrupole moments of two point redshift space correlation functions as data to fit, using our knowledge of symmetries within the model. We start with the covariance matrix and mean of the the multipoles measured from 1,000 Quijote mocks \cite{quijotesims}. We then add scatter by scaling the covariance matrix to a volume of a 10 ($h^{-1}$Gpc)$^3$ box.\footnote{Ignoring super-sample covariance \cite{howlett_percival_2019_supersamplecov}.} We apply a standard Markov-chain Monte Carlo (MCMC) analysis using the \texttt{emcee} python package \cite{emcee} and the following posterior distribution \cite{percival2021}:

\begin{equation}
    f(\xi^{\text{model}}|\xi^{\text{data}}, C) \propto \left[ 1+\frac{\chi^2}{n_s -1}\right]^{-\frac{m}{2}}, 
\end{equation}
where $n_s$ is the number of mocks used and $m$ is given by Equation 54 in \cite{percival2021}. $\chi^2$ is the given by $(\xi^{\text{model}}-\xi^{\text{data}})^TC^{-1}(\xi^{\text{model}}-\xi^{\text{data}})$ with $C^{-1}$ the inverse of the estimated covariance matrix:

\begin{equation}
    C_{ij}[\xi(r_{i})\xi(r_{j})] = \frac{1}{n_s(n_s - 1)} \sum^{n_s}_{n=1} [\xi_{n}(r_{i}) - \bar{\xi}(r_{i})][\xi_{n}(r_{j}) - \bar{\xi}(r_{j})]
\end{equation}
evaluated for all points in $\xi=\xi^{\text{data}}=\xi_{cc} - 2\xi_{c_nc_f}$. The BAO fit is performed using the monopole and quadrupole within the range $r\in [50,150]~h^{-1}$Mpc.

\subsection{The effect of interlopers on clustering measurements}  \label{sec:interlopers}

In general, when measuring redshifts, it is ideal to detect multiple emission lines per source to increase the confidence in the measurement. For this reason, the Nancy Grace Roman Space Telescope (Roman) galaxy redshift survey has mandated detections of at least two lines for all redshift measurements for its primary survey \cite{wang2022}. However, to increase the survey volume and provide a larger sample size, it is possible to determine redshifts when only a single line is measured. In this case, the possibility of line confusion now increases, as the assumed rest frame wavelength may not same as the truth. This can lead to galaxies with strong H$\beta$ emissions being mistaken for [OIII] sources. This causes the measured redshift to be incorrect, and as distances are derived from redshifts, so are the galaxy positions. This shift in position is unfortunately close to the BAO scale for [OIII]-H$\beta$ interlopers and heavily skews BAO measurements \cite{massara2020}. Calibrating this systematic is critical, else the improved statistical power is outweighed by interloper contamination. Several methods for this calibration have been introduced. \cite{foroozan2022, nguyen2024} have introduced modifications to the standard BAO template, allowing for the simultaneous modelling of the interloper contamination and the cosmological information. \cite{massara2023} introduced a machine learning method to constrain the fraction of galaxies in a catalogue that are interlopers.

The interloper displacement is only a function of the offset between the assumed and true rest frame wavelengths of the observed emission line, and the fiducial cosmology used to convert redshifts to distances. In the case of [OIII]-H$\beta$ interlopers, the displacement is given by:
\begin{equation}
    \Delta_{\rm fid} \approx 87.41 \frac{1 + z_{\text{true}}}{(\Omega_{\Lambda,{\rm fid}} + \Omega_{m,{\rm fid}}(1 + z_{\text{true}})^{3})^{1/2}} \, h^{-1}\text{Mpc}\,,
\end{equation}
with $\Omega_{\Lambda,{\rm fid}}$ and $\Omega_{m,{\rm fid}}$ being the energy density of the cosmological constant and matter in the fiducial cosmology, respectively. This displacement ranges from $80\,h^{-1}$Mpc to $97\,h^{-1}$Mpc, over the redshifts that Roman will survey.

With the introduction of interlopers to the galaxy sample, the catalogue becomes contaminated. A fraction, $f_i$, of objects are displaced by the above offset along the line of sight. We denote the full contaminated correlation function $\xi_{cc}$ and the contaminated overdensity $\delta_c$. We denote to the populations of interlopers and galaxy targets with the subscripts ``$i$" and ``$g$" respectively. With this notation, we write the overdensity,
\begin{equation}
    \delta_{c}(\Vec{x}) = (1 - f_{i}) \delta_{g}(\Vec{x}) + f_{i} \delta_{i}(\Vec{x})\,,
    \label{eq:overdensity}
\end{equation}
and the correlation function,
\begin{align}
    \xi_{cc}(\Vec{r}, f_{i})
    &= (1 - f_i)^{2} \xi_{gg}(\Vec{r}) + f_{i}^{2} \xi_{ii}(\Vec{r}) + 2f_i(1-f_i) \xi_{gi}(\Vec{r})\,,
    \label{eq:cc_corr}
\end{align}
with $\Vec{r} = \Vec{x_2} - \Vec{x_1}$. We have used the convention of \cite{foroozan2022}, that is the interlopers are considered at their observed positions. There are two types of components here, the auto correlations of the galaxies and interlopers, containing the BAO feature, and the galaxy-interloper cross correlation. It is this cross-term that skews the BAO in the correlation. Consider the interlopers at their true positions. They show a strong small scale clustering with the target galaxies. When they are shifted along the line of sight, this signal is incorrectly shifted to the displacement scale.

We test the power of LOO-PIT to detect interlopers using 1,000 halo mocks from the  Quijote suite of N-body simulations \cite{quijotesims}. We prepare the catalogues following the method of \cite{nguyen2024}. Here we briefly overview the procedure. See the aformentioned paper for a thorough discussion.

For each halo catalogue, we select a fraction $f_i$ of the halos to be interlopers and shift them towards the observer along the line of sight by a given displacement $\Delta$. We generate catalogues with interloper fractions $f_i$ = [0.005, 0.01, 0.02, 0.05, 0.10, 0.15]. We also consider several interloper shifts $\Delta$ = [85, 90, 97] ${h}^{-1}\rm Mpc$ to demonstrate the dependency of the statistic on the nature of the contamination. In particular, we consider the dependency of the statistic for contamination that introduces a second peak $\Delta$ = $85\,{h}^{-1}\rm Mpc$ compared to contamination that distorts the BAO peak $\Delta$ = [90, 97] ${h}^{-1}\rm Mpc$.

\begin{figure}[htbp]
\centering
\includegraphics[width=\textwidth]{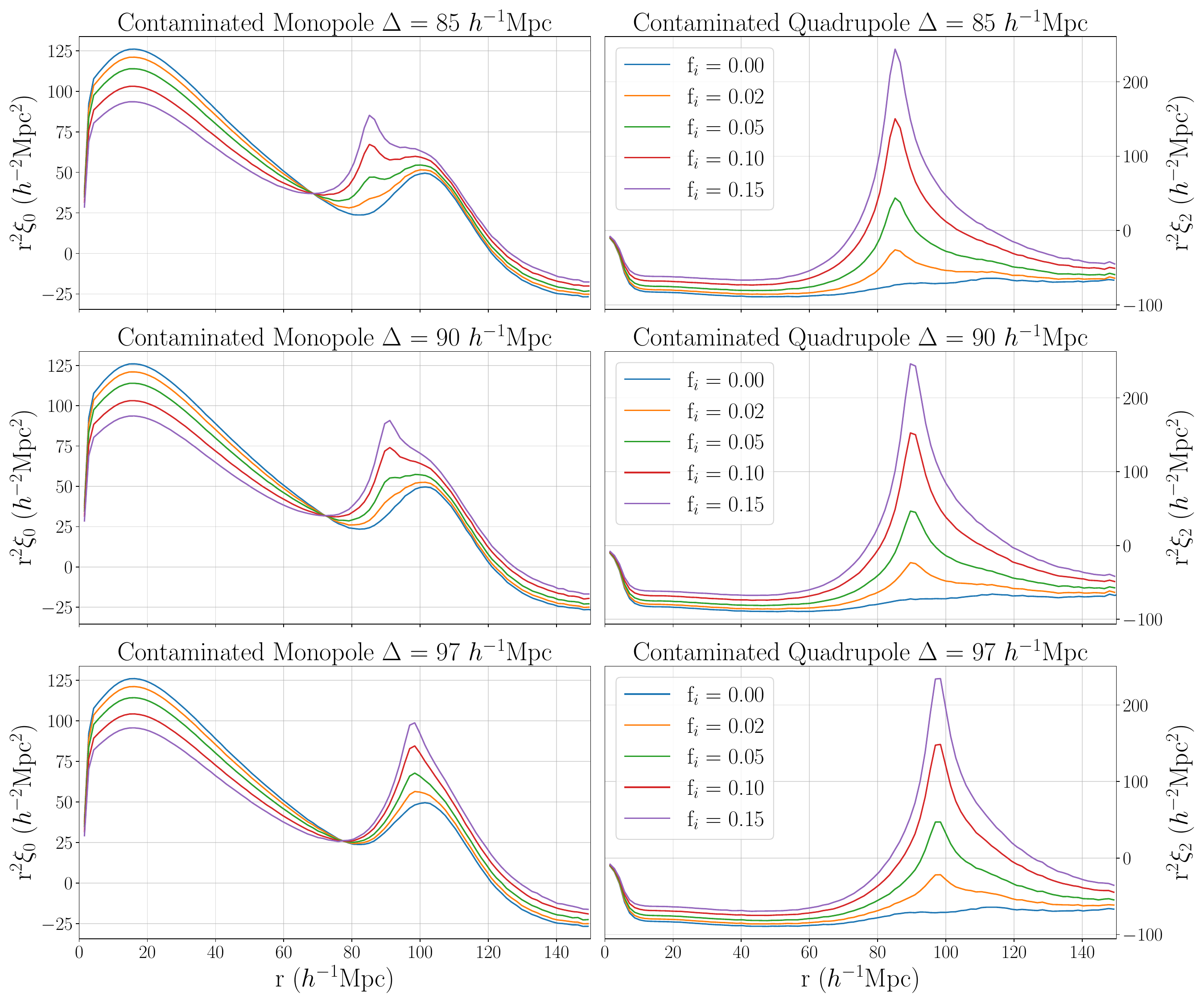}
\caption{The contaminated correlation function multipoles over a range of interloper fractions and displacements. \label{fig:contaminated_corrfunc}}
\end{figure}

We measure the correlation function using Nbodykit \cite{nbodykit} and extract the monopole and quadrupole moments. The effect of interlopers on the redshift-space monopole and quadrupole are shown in Figure~\ref{fig:contaminated_corrfunc}, showing the average of 1,000 contaminated correlation function measurements with various interloper fractions. 

For the correlation function templates, we begin by obtaining the real space linear power spectrum estimate, $P_{\rm lin}$, with the fiducial cosmology of the Quijote simulations, from \texttt{CAMB} \cite{camb}. We then include non-linear BAO damping effects and redshift space distortions, ignoring the Fingers-of-God (FoG) effect as we are using halos.\footnote{See \cite{nguyen2024} for a more detailed exposition.} From here, it is trivial to transform the power spectra into correlation function templates for use in the BAO model.

\subsection{BAO modelling results}
\label{sec:bao_results}

\begin{figure}[htbp]
\centering
\includegraphics[width=0.8\textwidth]{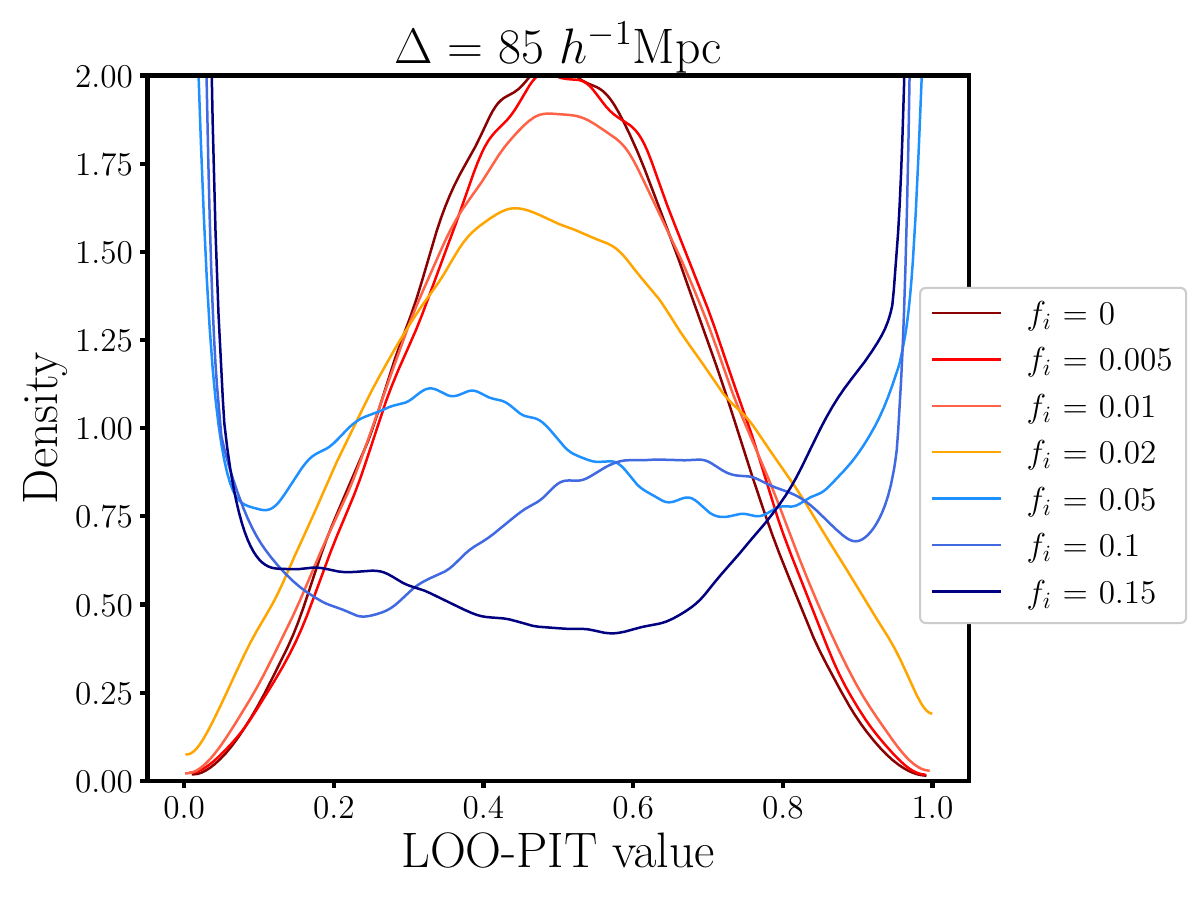}
\caption{LOO-PIT values averaged over 1,000 data realisations with $\Delta$ = 85 $h^{-1} {\rm Mpc}$. We plot a normalised KDE for each interloper fraction $f_i$. \label{fig:LOOPIT_85}}
\end{figure}

\begin{figure}[htbp]
\centering
\includegraphics[width=0.8\textwidth]{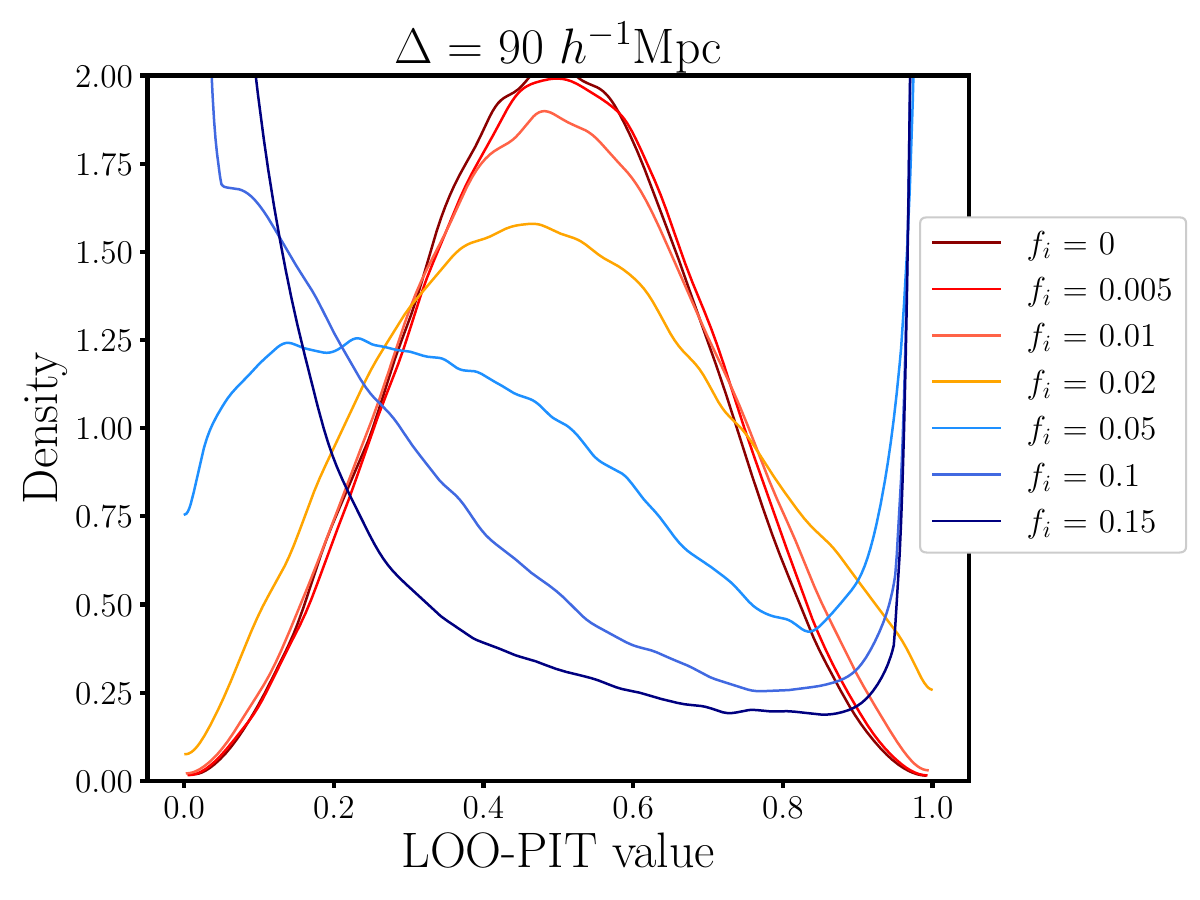}
\caption{LOO-PIT values averaged over 1,000 data realisations with $\Delta$ = 90 $h^{-1} {\rm Mpc}$. We plot a normalised KDE for each interloper fraction $f_i$. \label{fig:LOOPIT_90}}
\end{figure}

\begin{figure}[htbp]
\centering
\includegraphics[width=0.8\textwidth]{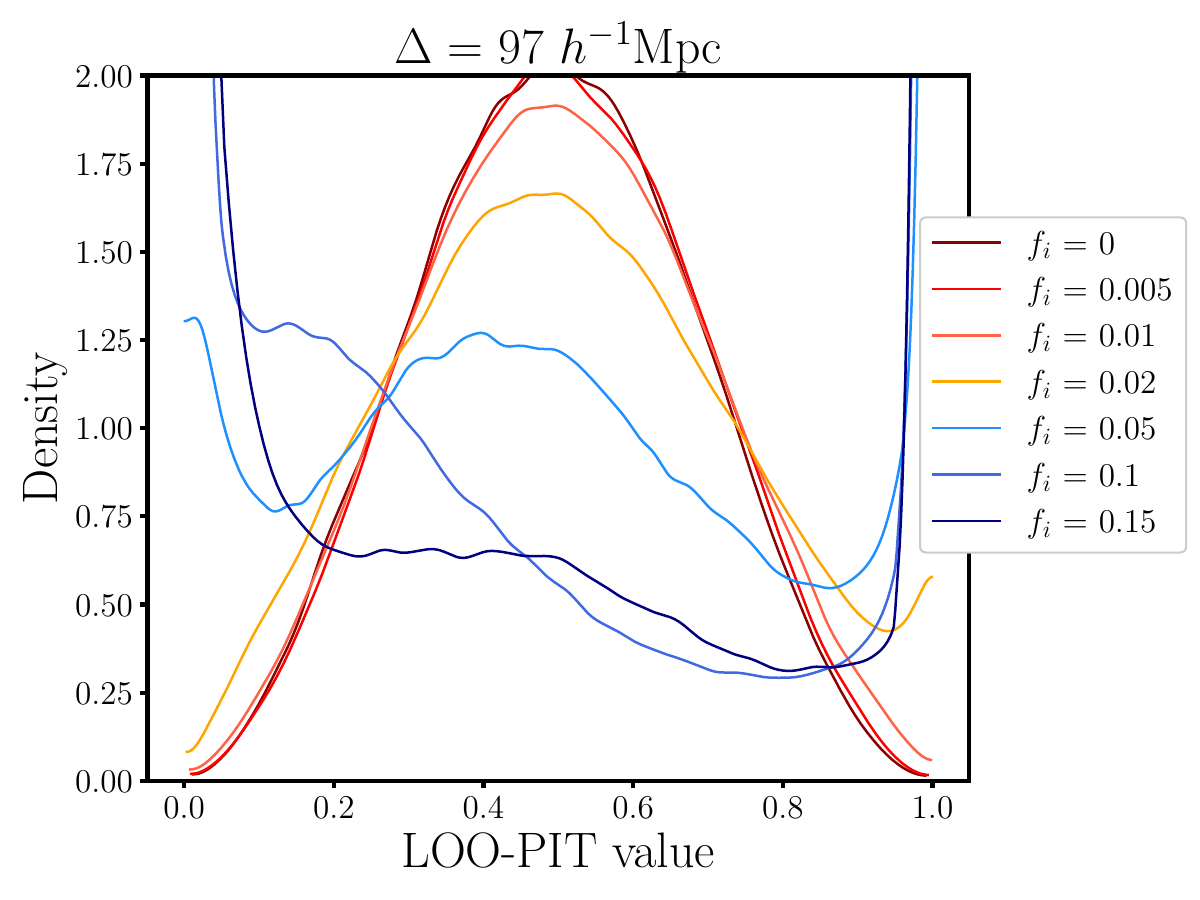}
\caption{LOO-PIT values averaged over 1,000 data realisations with $\Delta$ = 97 $h^{-1} {\rm Mpc}$. We plot a normalised KDE for each interloper fraction $f_i$. \label{fig:LOOPIT_97}}
\end{figure}

We perform the BAO fit on the monopole and quadrupole extracted from catalogues that contain no interlopers. For this baseline case, we generated 1,000 realisations of the data using the covariance matrix scaled to a volume of 10 ($h^{-1}$Gpc$)^3$, approximately matching the volume expected from Roman \cite{wang2022}. We then perform the BAO fit with the standard MCMC treatment, obtaining posterior distributions for all model parameters and use the chains to derive LOO-PIT draws according to the prescription outlined in Section~\ref{sec:method_outline}. This results in 132,000 total draws, and we plot the corresponding KDE in dark red in Figures~\ref{fig:LOOPIT_85},~\ref{fig:LOOPIT_90},~\ref{fig:LOOPIT_97}. We find that the baseline fit shows signs of significant overfitting with the characteristic ``peak'' shape (See Section~\ref{sec:common_signatures}).

We then consider catalogues with varying degrees of contamination. In this case, we generate 1,000 realisations of each combination of interloper fraction ($f_i$ = [0.005, 0.01, 0.02, 0.05, 0.10, 0.15]) and interloper displacement ($\Delta$ = [85, 90, 97] ${h}^{-1}\rm Mpc$), measure the two point correlation function monopole and quadrupole, and again obtain the LOO-PIT values for each case. We plot the KDE of these sets of 132,000 points in Figures~\ref{fig:LOOPIT_85},~\ref{fig:LOOPIT_90},~\ref{fig:LOOPIT_97}. We also show these plots with a single realisation in Appendix~\ref{appen:figures}, as these noisier distributions are closer to what a user will encounter. While the general trend is still visible without stacking, it is clear that there is some significant variance.

While the initial zero interloper baseline is non-uniform, LOO-PIT nevertheless is clearly sensitive to deviations from the baseline distribution. In general, the LOO-PIT plots start with a ``peak" shape, indicating overfitting. As the contamination increases, the height of the peak decreases. The distribution flips to the $trough$ shape between interloper fractions $f_i$ = 0.02-0.05. In particular, we can see a transition shape at $f_i$ = 0.05. There, we see both a bump and the underfitting/underdispersion signature, forming an offset ``$W$" shape. This is most clearly shown in the $f_i$ = 0.05 case in Figure~\ref{fig:LOOPIT_97}.

We can also see some separation between the $\Delta$ = 85 $h^{-1} {\rm Mpc}$ (Figure~\ref{fig:LOOPIT_85}) case compared to the other two. While, the LOO-PIT KDE deviates in a similar way for all three cases up until $f_i$ = 0.02, at $f_i$ = 0.05, there is a significant negative tilt present in the $\Delta$ = [90, 97] $h^{-1} {\rm Mpc}$ plots (Figures~\ref{fig:LOOPIT_90},~\ref{fig:LOOPIT_97}). Conversely, there is no tilt for $\Delta$ = 85 $h^{-1} {\rm Mpc}$. These are more clearly seen in the cases with greater levels of contamination. This suggests systematic bias in the fit. A tilt in the negative direction seen in the plots indicates the model is generally underestimating the data. For the cases where the interloper peak has significant overlap with the BAO peak, the model tends to underestimate the interloper peak. In this regime, the interloper peak distorts the BAO in such a way that the standard shape is a reasonable fit, leaving the interloper contribution as outliers. However, as the interloper peak moves further from the BAO peak, these outlier points cannot be ignored by the model as their likelihoods are extremely small when compared to a standard BAO shape, and thus the model will tend to``average through'' the peak in this case, resulting in more uniform central LOO-PIT draws for the $\Delta$ = 85 $h^{-1} {\rm Mpc}$ case for higher interloper fractions.

\begin{figure}[htbp]
\centering
\includegraphics[width=\textwidth]{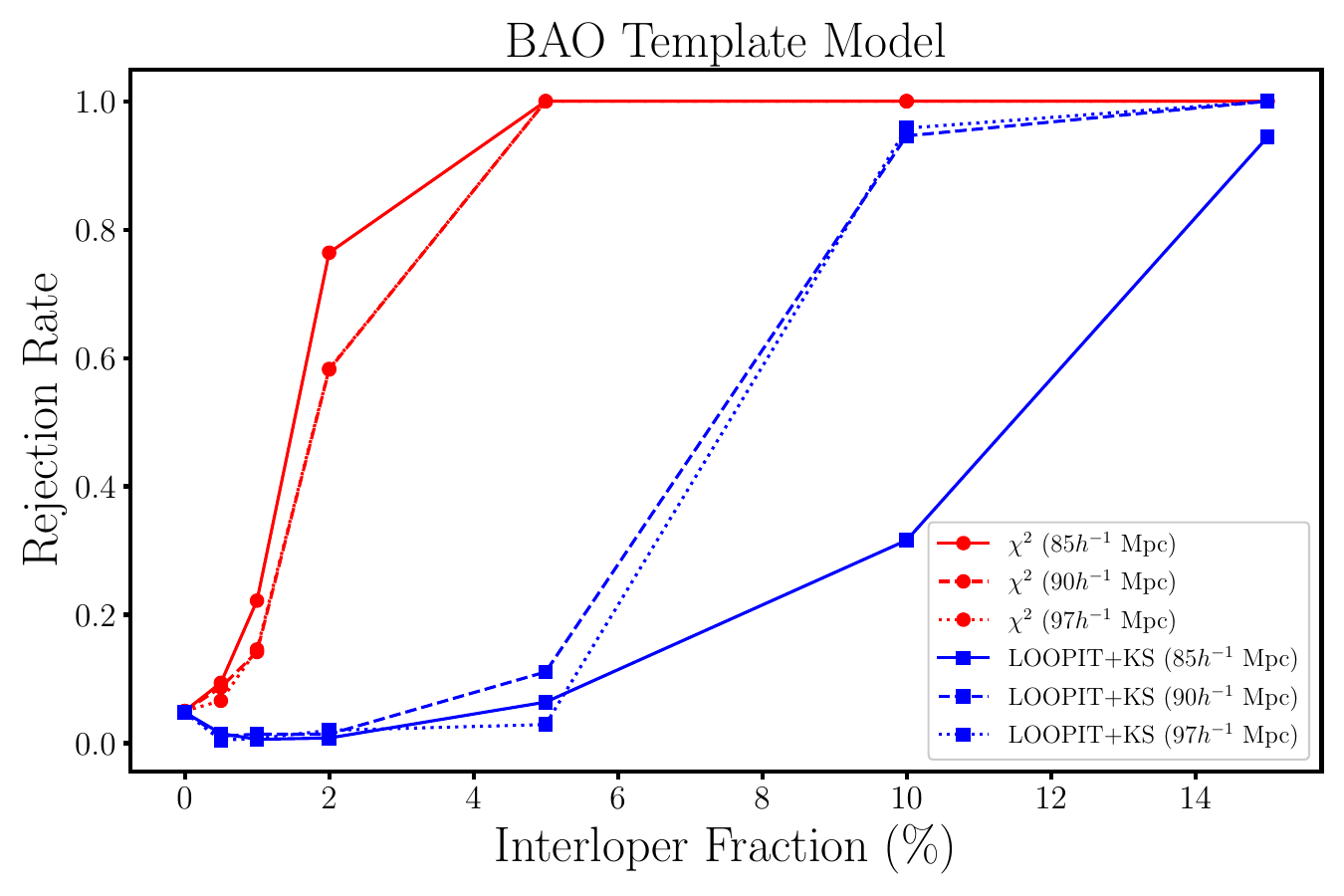}
\caption{Hypothesis test rejection rates as a function of interloper fraction for three interloper displacements. LOO-PIT+KS is plotted in blue. $\chi^2$ is plotted in red. \label{fig:chi2vsloopit}}
\end{figure}

We will now quantify the performance of LOO-PIT based tests compared to $\chi^2$. We consider the 132,000 LOO-PIT draws for the uncontaminated fit sufficient to  characterise the baseline distribution. This allows us to correctly control the Type I error rate, as well as marginalise over the correlation between individual LOO-PIT draws.
We compute the two sample Kolmogorov--Smirnov statistic between LOO-PIT draws with contamination and the baseline draws. We obtain the KS statistic from this and derive a p-value for each realisation by comparing to the empirical Kolmogorov distribution. Similarly, we keep the best fit $\chi^2$ statistic from each MCMC run and calculate a p-value with 123 (132 observations -- 9 parameters) degrees of freedom. We set a significance threshold at $\alpha$ = 0.05, and consider the fraction of realisations that result in a rejection of the null for LOO-PIT and then $\chi^2$. We plot these in Figure~\ref{fig:chi2vsloopit}.

As expected, we see similar behaviour to the simple model test cases (see the upper panel of Figure~\ref{fig:toy_model}). For this type of sharply peaked contamination feature, $\chi^2$ is a more powerful test overall. However, note the difference between the $\Delta$ = 85 $h^{-1} {\rm Mpc}$ and $\Delta$ = [90, 97] $h^{-1} {\rm Mpc}$ cases: in general, LOO-PIT gains power in the latter cases. These are also the same cases where the tilt signature is present in the LOO-PIT KDE plots. This is in line with our expectations. LOO-PIT becomes a more powerful test (relative to the $\chi^2$) the more the contamination is similar to a small bias, such as the uniform shift as shown in the lower panel of Figure~\ref{fig:toy_model}.\footnote{We could imagine a scale from a single outlier to a Gaussian feature gradually widening until the bias is uniform across the data.} We also note the dip in the rejection rate at low interloper fraction. This is due to the relative shift in the estimated Kolmogorov distributions for low contamination. We show an example of these in Figure~\ref{fig:k_dists}.

\subsection{LOO-PIT for larger displacement interlopers}
\label{sec:large_displacement_loopit}

Here we consider a scenario where LOO-PIT is more powerful. In general, we expect a range of possible interloper displacements. For example, larger scale interlopers are also likely to appear in the galaxy catalogues. Their effect on the modelling is not as large as small displacement interlopers close to the BAO peak. Here, we consider applying LOO-PIT to catalogues with $f_i$ = [0.02, 0.05, 0.10, 0.15] and $\Delta$ = 170 ${h}^{-1}\rm Mpc$. In this case, the galaxy-interloper peak is outside of the fitting range and its effect is to add in a small scale dependent bias across the fitting range. This is similar to the uniform bias considered in the toy model, however the slight change in amplitude of this bias across the fitting range prevents the polynomial terms to fully capture it.

Here we also introduce a new statistic that is slightly more powerful than KS in this case. This is the sum-of-squares statistic (SS):
\begin{equation}
    {\rm SS} = \sum_{i} \left({\rm ECDF_1}(i) - {\rm ECDF_2}(i)\right)^2\,,
\end{equation}
where we consider the sum of squared differences between the empirical cumulative distribution functions (ECDF) of the two distributions being compared. We sum over $i$, where $i$ indexes the quantile where we evaluate the ECDFs. While KS only considers the supremum of the set of absolute difference between the ECDF, by considering a sum over all the differences, we gain power in detecting systematic differences, as opposed to being more sensitive to outliers. We again calculate an empirical null distribution of the SS statistic and proceed as usual with the hypothesis testing.\footnote{We also applied the SS statistic to the modelling in Section~\ref{sec:bao_results}, but the change in the conclusions were negligible. We have included a version of Figure~\ref{fig:chi2vsloopit} using the SS statistic in Appendix~\ref{appen:figures}.}

\begin{figure}[htbp]
\centering
\includegraphics[width=\textwidth]{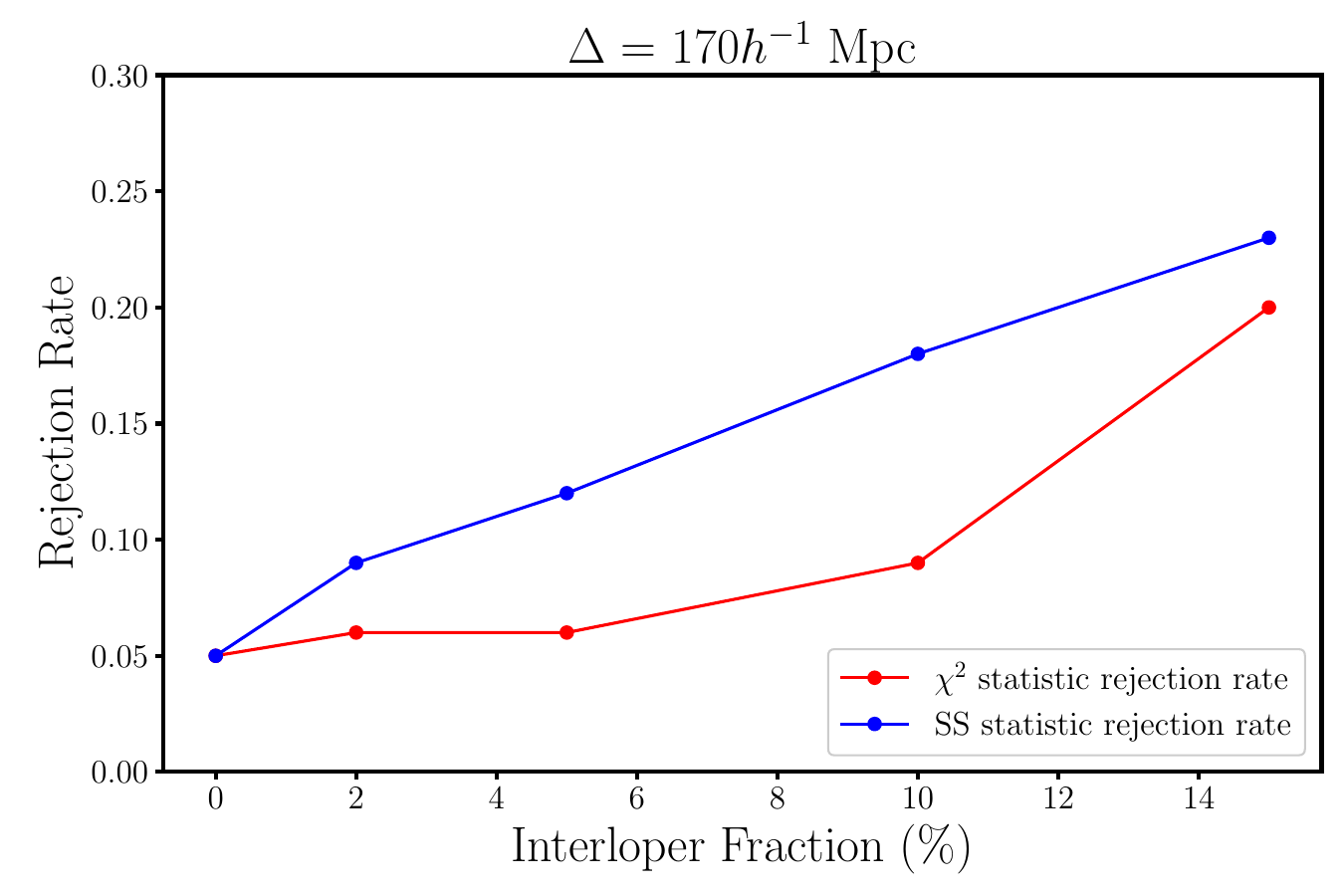}
\caption{Hypothesis test rejection rates as a function of interloper fraction for a large interloper displacement. LOO-PIT+SS is plotted in blue. $\chi^2$ is plotted in red. Note the re-scaled vertical axis. \label{fig:chi2vsloopit_ld}}
\end{figure}

Similar to the above analyses, we consider 100 realisations of the data for each interloper fraction and consider the power of each statistic to reject the null hypothesis. We show this in Figure~\ref{fig:chi2vsloopit_ld}.

Here we have shown a scenario where LOO-PIT+SS outperforms $\chi^2$, due the type of contamination. We also see that the detections regardless of method are low, as the effect of these large displacement interlopers are small. In general, $\chi^2$ tends to be more powerful when detecting spurious outliers, while LOO-PIT tends to have more power when detecting systematic offsets or biases between the data and model. This would imply that LOO-PIT and $\chi^2$ would reject different realisations, and this is indeed the case. For these large displacement interloper realisations (not including the null cases), LOO-PIT+SS rejects 57, $\chi^2$ rejects 36, and both reject 5. Thus out of a total of 400 realisations, we have a total of 88 rejections and the overlap between the detection methods is small.\footnote{There are 4 interloper fractions, and 100 realisations for each.} Thus, the intersection of these rejections is small and it would be beneficial to use both these tests in conjunction to improve our power in detecting modelling issues like interlopers. 

\section{Calibration of the baseline}
\label{sec:baseline_calibration}

The core idea of LOO-PIT is that a correctly calibrated model will yield a uniform LOO-PIT KDE. Indeed, that was the standard approach outlined in Section~\ref{sec:method_outline} and the case for the toy model fits in Section~\ref{sec:simp_model}. However, as the BAO template model is designed to extract the AP parameters from the correlation function, it intentionally overfits the broadband power with the polynomial terms. This limits the ability of the standard LOO-PIT approach to detect the difference between an uncontaminated and a contaminated correlation function. We cannot compare the LOO-PIT KDE directly to the uniform distribution, as it is not the correct null distribution in this case. To counter this, we extract LOO-PIT samples from data that has no contaminants, to characterise the null distribution for this test case. This allows us to visually see the difference between the null LOO-PIT distribution and any possible deviations and allows distribution tests to determine the significance of the detection of contaminants.

Above, we have compared draws of the null and contaminated LOO-PIT samples to produce the power plots. To correctly calibrate the test, we had to estimate the Kolmogorov (or SS) distribution using our large set of null draws. While the Kolmogorov distribution is well known and has tables of critical values, we found that using the standard values would result in a high Type I error rate. These critical values are dependent only on the sizes of the samples being compared, assuming each element of each sample is statistically independent. This is not the case for our LOO-PIT draws. As they are dependent on the covariance matrix of the data, they inherit some of the correlation of the data points. Indeed, the covariance matrix of our BAO data is far from sparse, thus the LOO-PIT draws themselves are highly correlated. By comparing to the standard Kolmogorov distribution tables, we find that our empirical null distribution is consistent with samples of effective size $n\sim$30.

Our example shows that an estimate of a baseline LOO-PIT distribution must be done before any comparisons can be made, especially in the case where one wishes to detect problems using an hypothesis test. This approach will be straightforward for the detection of anomalies in cosmological data, as it is now commonplace to validate data pipelines on sophisticated simulations even prior to the beginning of survey missions. In these cases, it is a non-issue to calibrate a baseline distribution on a set of simulations under a fiducial cosmology, as we have done in this paper. 

For our example, it would be beneficial to run LOO-PIT on a set of Roman-like mocks in the standard  $\Lambda$CDM cosmology in the absence of observational systematics. Regardless of the model used for BAO or any other observation in these simulations, we could characterise the LOO-PIT distribution for a ``clean" detection, allowing for us to detect deviations, like observational systematics and their severity or new physics, from this assumption as real data  from Roman becomes available. 

We should note that this extra step is only relevant in the case where a researcher may wish to detect an anomaly at some significance, or to compare a series of possible models. Of course, no baseline is required to perform the LOO-PIT procedure, only to perform a distribution test to determine detection significance. Thus, LOO-PIT may also be useful in the model building stage of the pipeline. The standard peak and trough signatures are useful for calibrating error analyses like covariances, and building a model that returns the uniform can validate a model as well calibrated.

\section{Conclusions}
\label{sec:conclusions}

In this paper, we have described a  statistical method called the Leave One Out-Probability Integral Transform (LOO-PIT) and discussed its use within astronomy. This is a general technique with a wide variety of applications. We demonstrated LOO-PIT using the problem of detecting small displacement interlopers in galaxy clustering analyses.

LOO-PIT combines two statistical techniques together: Leave One Out-Cross Validation (LOO-CV), and the well known Probability Integral Transform (PIT). We estimate a LOO posterior predictive distribution for each data point, then test this estimate via the PIT. Our test diagnostic is the KDE of LOO-PIT values, one for each data point. We can also quantify the diagnostic by applying hypothesis testing via either the KS or SS statistic.

We tested the effectiveness of LOO-PIT by applying it to the analysis of contaminated correlation functions. We measured the mean of 1,000 monopole and quadrupole moments from simulations, as well as the covariance matrices. We then scale the covariance matrix to a Roman like volume of 10 ($h^{-1}$Gpc)$^3$ box. We generate 1,000 realisations of multipoles at this volume using the scaled covariance matrix. We fit contaminated correlation functions with various values of the interloper fraction and displacement. We find that the shape of the LOO-PIT KDEs are change significantly between $f_i$ = 0.02-0.05 for interloper displacements $\Delta$ = [85, 90, 97] $h^{-1} {\rm Mpc}$ . At this threshold, it is clear by eye to see deviation from the baseline. We also test LOO-PIT at an interloper displacement $\Delta$ = 170 $h^{-1} {\rm Mpc}$, mimicking a small scale bias across the data. In this case, LOO-PIT+SS is a more powerful test than $\chi^2$ for all interloper fractions considered.

LOO-PIT has several primary advantages over the simple $\chi^2$ test. First, whereas the $\chi^2$ test relies on a Gaussian assumption, LOO-PIT does not. Second, while $\chi^2$ uses information from a single point in the parameter space, LOO-PIT can incorporate information from the entire posterior distribution. This can lead to a better global assessment of the inference, such as in assessing the calibration of priors \cite{hod_priors, sim_priors}. Lastly, the strongest advantage of using LOO-PIT is that it includes more information about the type of model failure. While LOO-PIT detects the same overdispersion and underdispersion as $\chi^2$, LOO-PIT can detect biases in the data relative to the model that $\chi^2$ is blind to. This is visible in the LOO-PIT KDE diagnostic plots. This additional information that LOO-PIT provides makes it a complementary tool to the current methods for model comparison, and detection of systematics. We find that LOO-PIT can provides power statistical power in detecting certain biases that $\chi^2$ cannot. This difference in detection provides a significant incentive to use LOO-PIT and $\chi^2$ in conjunction, giving the combined test greatly improved power compared to either LOO-PIT or $\chi^2$.

\appendix

\section{LOO-PIT without stacking}
\begin{figure}[htbp]
\centering
\includegraphics[width=\textwidth]{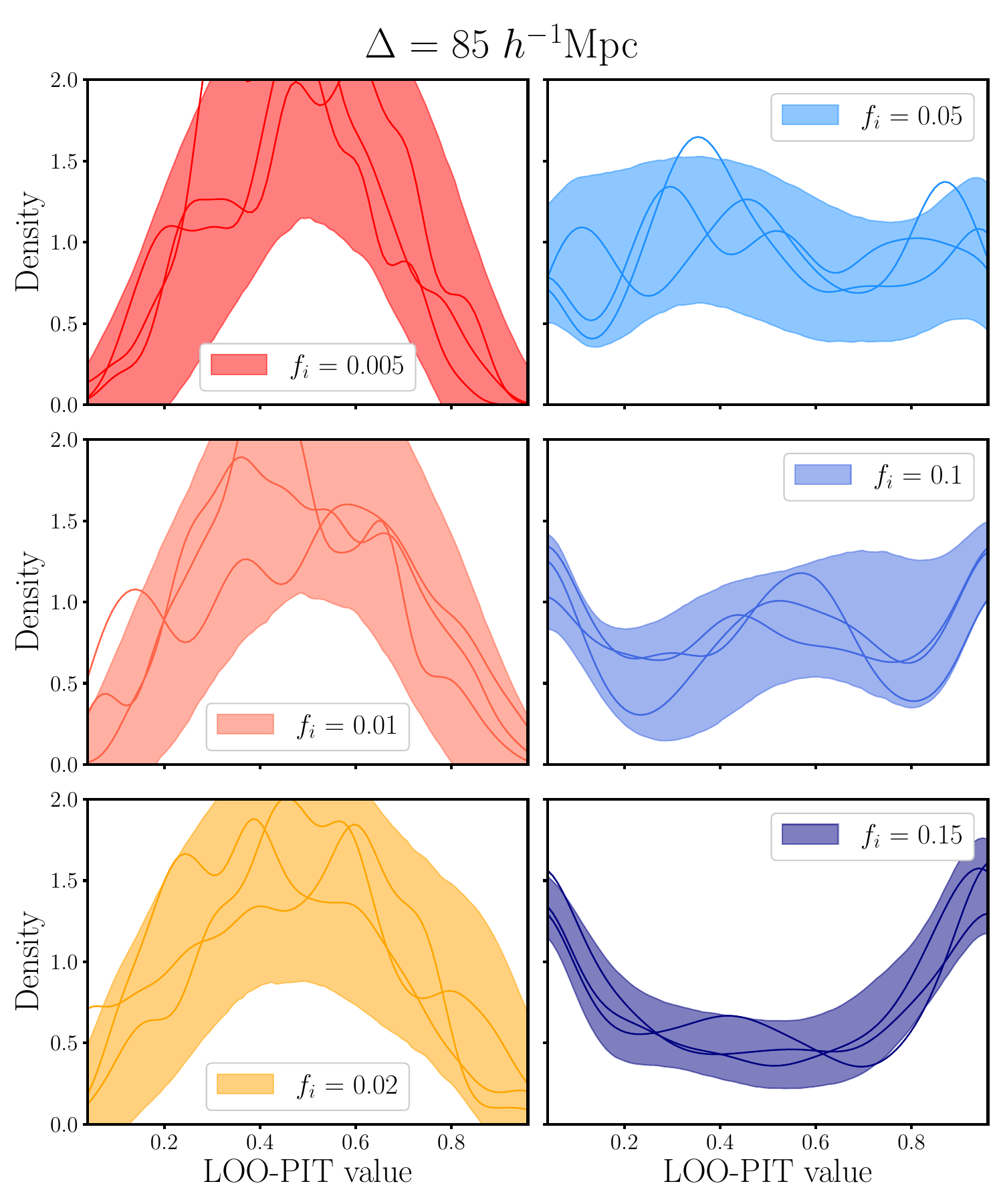}
\caption{LOO-PIT values for $\Delta$ = 85 $h^{-1} {\rm Mpc}$. We plot $2\sigma$ confidence bands for each interloper fraction. We also overplot three LOO-PIT KDEs, each derived from a single realisation. \label{fig:LOOPIT_85_ribbon}}
\end{figure}

\begin{figure}[htbp]
\centering
\includegraphics[width=\textwidth]{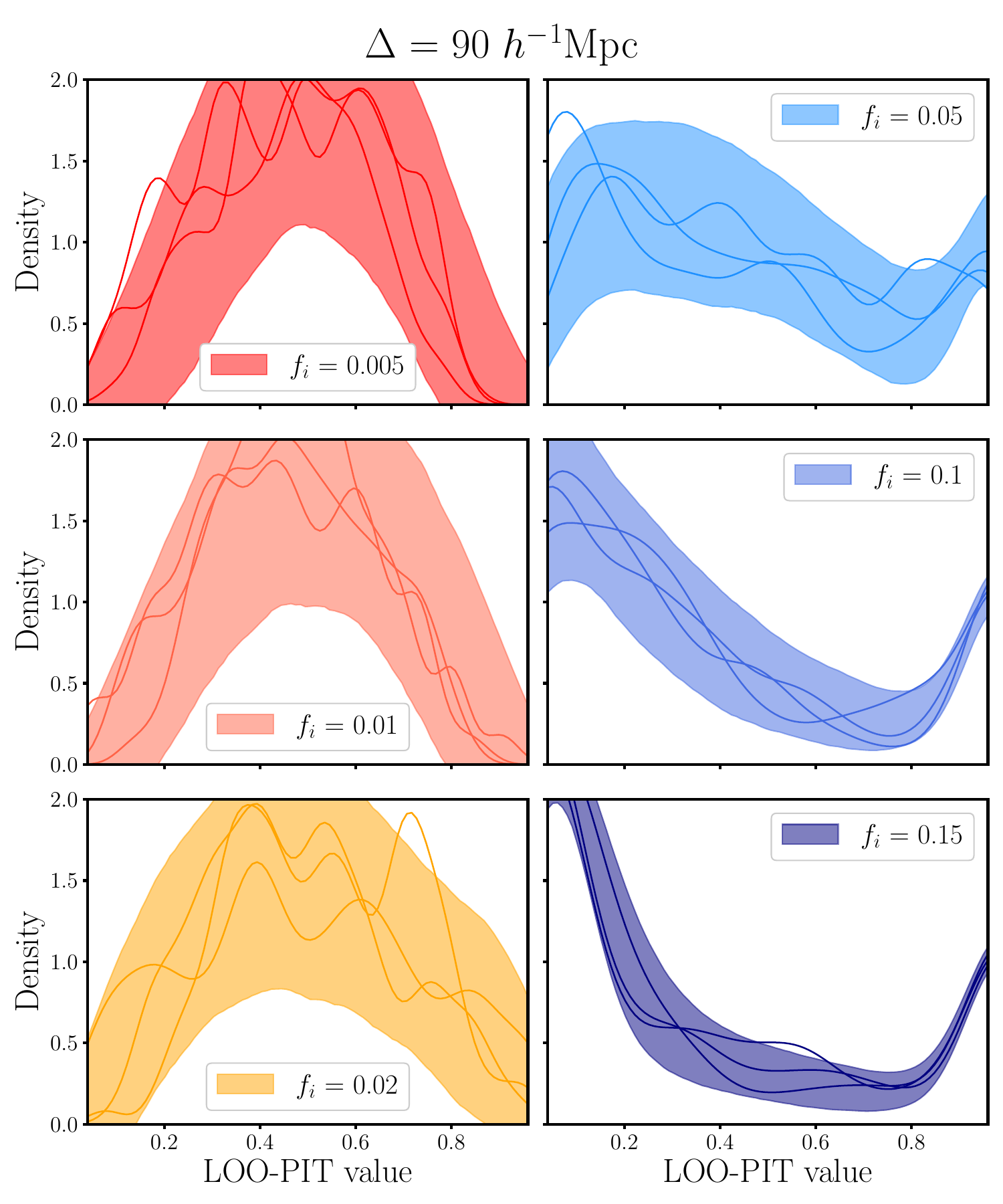}
\caption{LOO-PIT values for $\Delta$ = 90 $h^{-1} {\rm Mpc}$. We plot $2\sigma$ confidence bands for each interloper fraction. We also overplot three LOO-PIT KDEs, each derived from a single realisation. \label{fig:LOOPIT_90_ribbon}}
\end{figure}

\begin{figure}[htbp]
\centering
\includegraphics[width=\textwidth]{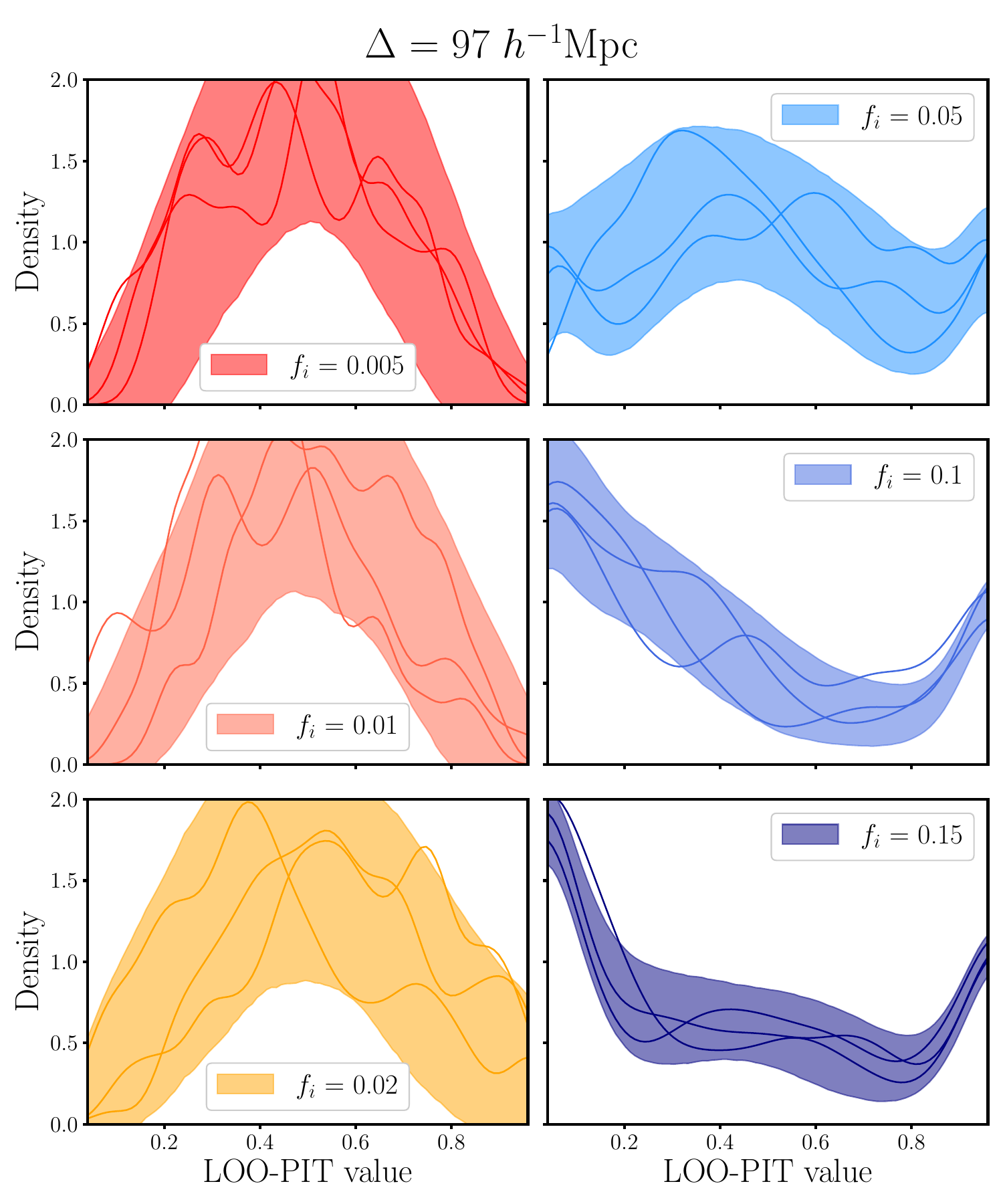}
\caption{LOO-PIT values for $\Delta$ = 97 $h^{-1} {\rm Mpc}$. We plot $2\sigma$ confidence bands for each interloper fraction. We also overplot three LOO-PIT KDEs, each derived from a single realisation. \label{fig:LOOPIT_97_ribbon}}
\end{figure}

In Figures~\ref{fig:LOOPIT_85}, \ref{fig:LOOPIT_90} \& \ref{fig:LOOPIT_97} we presented results for the visual LOO-PIT plots by stacking realisations to show smoother plots. Because of the complexity of these plots it was not possible to indicate errors for the results. As with all distribution tests, the accuracy of the test increases with the size of the offset compared to the statistical errors. 
To understand this, we shown examples from individual simulations in Figures~\ref{fig:LOOPIT_85_ribbon}, \ref{fig:LOOPIT_90_ribbon} \& \ref{fig:LOOPIT_97_ribbon}. Clearly, there is some significant noise compared to the stacked plots. However, we note that the general trends described in Section~\ref{sec:bao_model} are still visible. In particular, the flip in shape between $f_i$ = 0.02-0.05 is still clear, as is the negative tilt at the higher interloper fractions. With a single realisation, LOO-PIT is more difficult to use to diagnose issues simply by inspection of the KDE plots. However, we will note that using $\chi^2$ yields no such visual inspection method.

\section{Additional Figures}
\label{appen:figures}

\begin{figure}[htbp]
\centering
\includegraphics[width=0.9\textwidth]{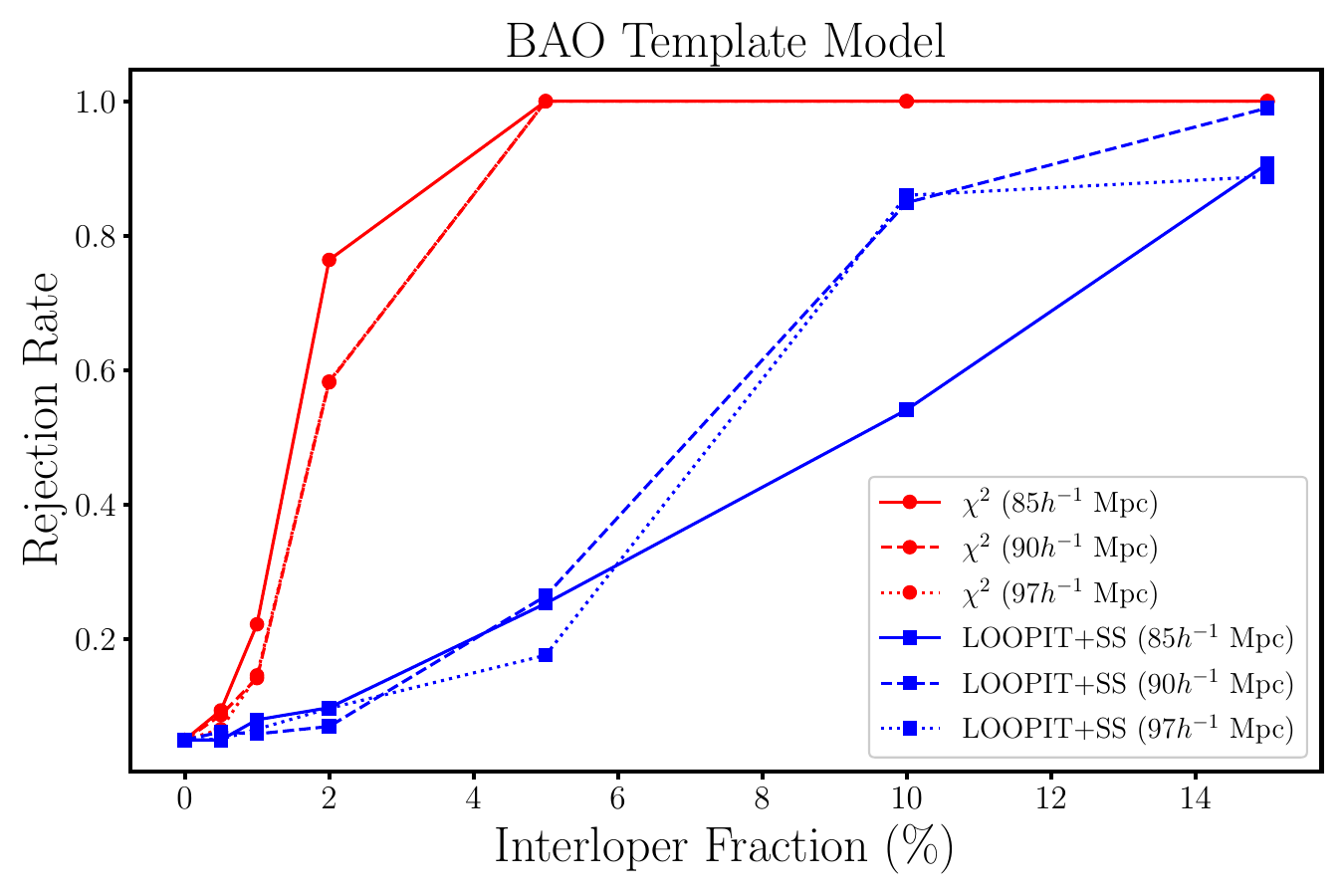}
\caption{Hypothesis test rejection rates as a function of interloper fraction for three interloper displacements. LOO-PIT+SS is plotted in blue. $\chi^2$ is plotted in red. \label{fig:chi2vsloopit_ss}}
\end{figure}

\begin{figure}[htbp]
\centering
\includegraphics[width=0.9\textwidth]{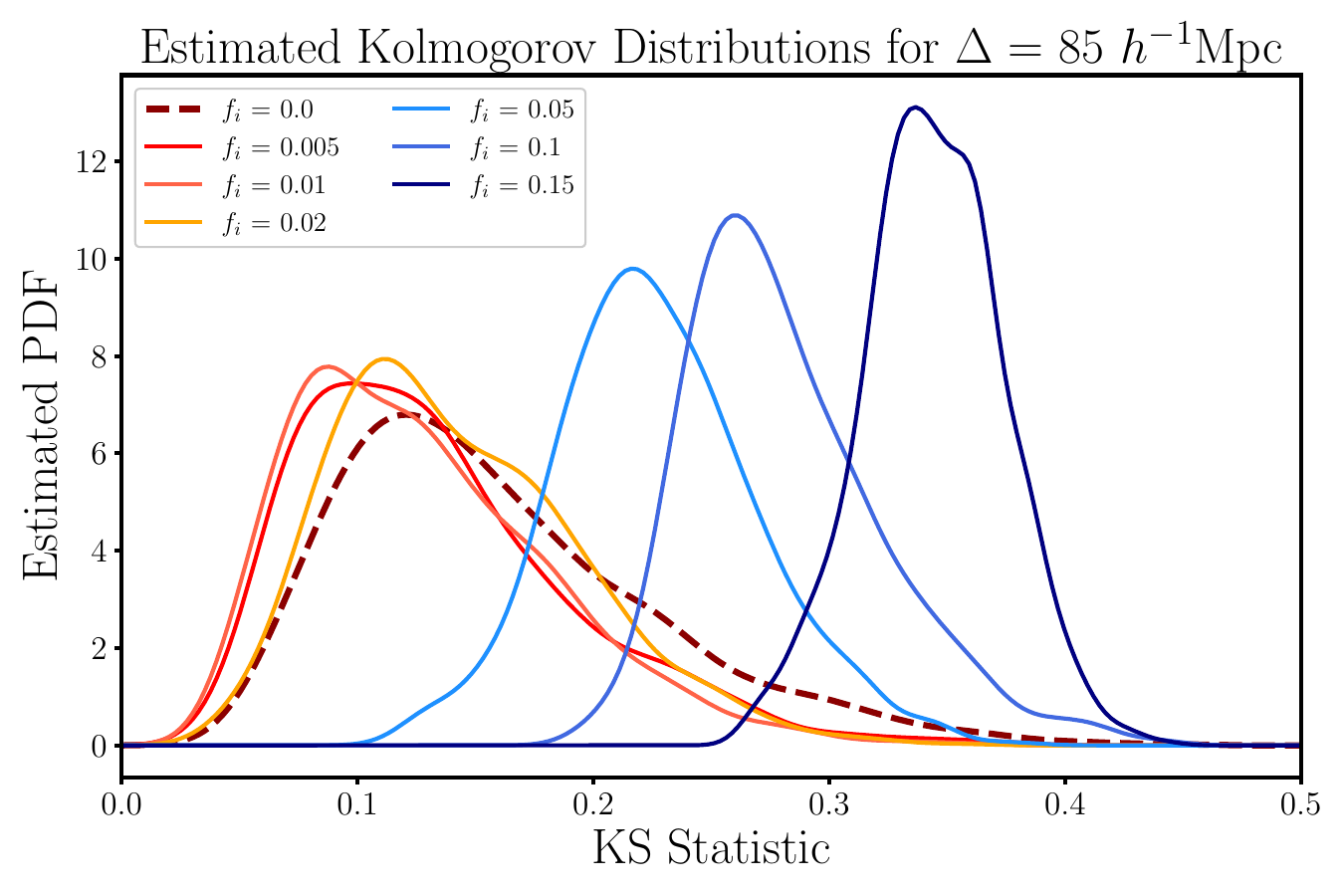}
\caption{Estimated Kolmogorov distributions for $f_i$ = [0.02, 0.05, 0.10, 0.15]. Note the difference in width between the null ($f_i$ = 0) and contaminated distributions. \label{fig:k_dists}}
\end{figure}

\acknowledgments

All authors acknowledge the support of the Canadian Space Agency. WP also acknowledges support from the Natural Sciences and Engineering Research Council of Canada (NSERC), [funding reference number RGPIN-2019-03908].

Research at Perimeter Institute is supported in part by the Government of Canada through the Department of Innovation, Science and Economic Development Canada and by the Province of Ontario through the Ministry of Colleges and Universities.

This research was enabled in part by support provided by Compute Ontario (computeontario.ca) and the Digital Research Alliance of Canada (alliancecan.ca). 

\printbibliography

\end{document}